\documentclass[reprint,aip,prl,amsmath,amssymb,reprint,twocolumn]{revtex4-2}

\usepackage{graphicx}                                   
\usepackage{dcolumn}                                    
\usepackage{bm}                                         
\usepackage{chemformula} 
\usepackage[utf8x]{inputenc} 
\usepackage[T1]{fontenc} 
\usepackage{mathrsfs}
\usepackage{braket}
\usepackage{amssymb}
\usepackage{appendix}
\usepackage{hyperref}
\usepackage{color}
\usepackage[normalem]{ulem}
\usepackage{dcolumn}
\usepackage{adjustbox}
\usepackage{tabularx}
\usepackage{tikz}
\usepackage{physics}
\usepackage{program}
\usepackage{graphicx}
\usepackage{booktabs}
\usepackage[normalem]{ulem}
\usetikzlibrary{arrows,shapes,positioning,shadows,trees}

\newcommand{\red}[1]{{\color{black} #1}}

\newcommand{\nl}{\nonumber \\}

\draft 

\begin{document}

\title{Machine learning meets $\mathfrak{su}(n)$ Lie algebra: Enhancing quantum dynamics learning with exact trace conservation} 
\author{Arif Ullah}   
\email{arif@ahu.edu.cn}
\affiliation{School of Physics and Optoelectronic Engineering, Anhui University, Hefei, 230601, Anhui, China}  
\author{Jeremy O. Richardson}   
\email{jeremy.richardson@phys.chem.ethz.ch}
\affiliation{Department of Chemistry and Applied Biosciences, ETH Z\"urich, 8093 Z\"urich, Switzerland}  

\date{\today}

\begin{abstract}
Machine learning (ML) has emerged as a promising tool for simulating quantum dissipative dynamics. However, existing methods often struggle to enforce key physical constraints, such as trace conservation, when modeling reduced density matrices (RDMs). While Physics-Informed Neural Networks (PINN) aim to address these challenges, they frequently fail to achieve full physical consistency. In this work, we introduce a novel approach that leverages the $\mathfrak{su}(n)$ Lie algebra to represent RDMs as a combination of an identity matrix and $n^2 - 1$ Hermitian, traceless, and 
orthogonal basis operators,
where $n$ is the system's dimension. By learning only the coefficients associated with the operators, our framework inherently ensures exact trace conservation, as the traceless nature of the operators restricts the trace contribution solely to the identity matrix. This eliminates the need for explicit trace-preserving penalty terms in the loss function, simplifying optimization and improving learning efficiency. 
We validate our approach on two benchmark quantum systems: the spin-boson model and the Fenna-Matthews-Olson complex. By comparing the performance of four neural network (NN) architectures—Purely Data-driven Physics-Uninformed Neural Networks (PUNN), \(\mathfrak{su}(n)\) Lie algebra-based PUNN (\(\mathfrak{su}(n)\)-PUNN), traditional PINN, and \(\mathfrak{su}(n)\) Lie algebra-based PINN (\(\mathfrak{su}(n)\)-PINN)—we highlight the limitations of conventional methods and demonstrate the superior accuracy, robustness, and efficiency of our approach in learning quantum dissipative dynamics.
\end{abstract}
 
\maketitle
\section{Introduction}
Open quantum systems provide a powerful framework for understanding the dynamics of quantum systems interacting with their environment. Such interactions are fundamental across various fields, including quantum information processing,\cite{breuer2016colloquium} quantum memory,\cite{khodjasteh2013designing} quantum transport,\cite{cui2006quantum} energy transfer in photosynthesis,\cite{zerah2021photosynthetic} and biochemical processes like proton tunneling in DNA.\cite{slocombe2022open} To describe these interactions, the total quantum state is typically divided into two components: the system of interest and its surrounding environment. The system's behavior is captured by the reduced density matrix (RDM), which evolves due to both the system’s intrinsic dynamics and the influence of the environment.


Modeling environmental effects within the system dynamics presents significant challenges due to the vast number of environmental degrees of freedom. Approaches to model these effects range from mixed quantum--classical treatments\cite{Miller2001SCIVR, cotton2013symmetrical, liu2021unified,runeson2019spin,runeson2020generalized, mannouch2020partially, mannouch2020comparison, mannouch2022partially, tao2016multi,mannouch2023mapping, crespo2018recent, qiu2022multilayer} to entirely quantum descriptions.\cite{ivander2024unified} In the mixed quantum--classical methods, the system is treated quantum mechanically, whereas the degrees of freedom of the environment, such as molecular vibrations, solvent motions, or other large-scale interactions, are represented by a classical phase space. 
Trajectories evolve according to classical equations of motion, typically derived from Newtonian or Hamiltonian mechanics. This significantly reduces the computational complexity because it bypasses the high-dimensional Hilbert space of the environment.

Fully quantum approaches, in contrast, treat both the system and its environment within the quantum Hilbert space, allowing for a complete description of quantum coherence, dissipation, and entanglement in the system’s dynamics. \red{These methods capture the full quantum nature of the interactions without relying on classical approximations. Path-integral approaches like the hierarchical equations of motion (HEOM)\cite{tanimura1989time, makarov1994path, su2023extended, yan2021efficient, gong2018quantum, xu2022taming, bai2024heom, wang2023simulating} and quasiadiabatic propagator path integral (QUAPI)\cite{makarov1994path, makri2023quantum} rigorously incorporate memory effects and non-Markovian behavior. Quantum master equation methods, including the stochastic equation of motion (SEOM)\cite{han2019stochastic, han2020stochastic, ullah2020stochastic, chen2022simulation, dan2022generalized, stockburger2016exact} and the generalized master equation\cite{lyu2023tensor,liu2018exact} provide alternative fully quantum descriptions of environmental interactions.}

Although both mixed quantum--classical and fully quantum methods provide a thorough description of open quantum systems, they come with their own set of challenges. Classical approaches, for example, may fail to capture 
detailed balance
\cite{Schmidt2008equilibrium,ellipsoid,thermalization}
or subtle quantum correlations.\cite{Mannouch2024coherence}
In contrast, while fully quantum methods offer a more accurate and complete representation of system-environment interactions, they often demand substantial computational resources, especially in scenarios involving strong system-environment coupling or when fine discretization steps are required to maintain numerical stability.

In recent years, machine learning (ML) has emerged as a powerful tool for learning complex spatio-temporal dynamics in high-dimensional spaces.\cite{ullah2021speeding, ullah2022predicting, ullah2022one, rodriguez2022comparative, herrera2021convolutional, ge2023four, zhang2023excited, wu2021forecasting, lin2022automatic, bandyopadhyay2018applications, yang2020applications,lin2022trajectory,tang2022fewest,  shakiba2024ml_spin_relaxation, lin2024enhancing, zeng2024sophisticated, long2024quantum, cao2024neural, zhang2024nonmarkov, zhang2024ai, herrera2024short,PhysRevResearch.7.L012013, ullah2025short} It has demonstrated proficiency in predicting the future evolution of quantum states based on historical data, as well as directly forecasting quantum states as a function of time and/or simulation parameters.

Despite the advantages of ML approaches, most methods directly learn and predict the dynamics of the individual RDM matrix elements%
, which often results in the violation of the fundamental physical principle of trace conservation. While physics-informed neural networks (PINN), which incorporate trace conservation into their loss function, help to mitigate trace violations, they still fail to fully preserve trace conservation, as demonstrated in our recent work.\cite{ullah2024pinn}

Taking inspiration from the success of the spin-mapping approach, 
\cite{runeson2020generalized} in this paper, we utilize the $\mathfrak{su}(n)$ Lie algebra to expand the RDM $\mathbf{\rho}_{\rm S}$ 
in terms of the identity operator and $n^2-1$ traceless operators. 
The key point is that for a properly normalized RDM, the coefficient of the identity operator remains constant under exact quantum-mechanical evolution.
We therefore only have to learn the $n^2-1$ time-dependent coefficients corresponding to the traceless operators.

This approach offers significant advantages over directly learning the dynamics of the RDM. By focusing on the coefficients associated with the traceless operators, we inherently eliminate the need to account for trace conservation in the loss function, as these operators are traceless and the identity operator explicitly ensures the correct trace. This guarantees that the reconstructed RDM will always have the correct trace, unlike approaches such as PINN, which often struggle to enforce trace conservation during training. This inherent trace preservation simplifies the loss function, reducing its complexity and thereby enhancing the learning process and model performance.



\section{Theory and Methodology}
Consider an open quantum system, denoted by $\rm S $, with $n$ states interacting with an external environment $\rm E $. The dynamics of the combined system $\rm S + E $ is described by the Liouville--von-Neumann equation (assuming $\hbar=1$):
\begin{equation}
    \dot{\mathbf{\rho}}(t) = -\mathrm{i}[\mathbf{H}, \mathbf{\rho}(t)],
\end{equation}
where $\mathbf{H}$ is the Hamiltonian and $\mathbf{\rho}(t)$ is the density matrix of the total system. Given that this system is closed, it follows a unitary evolution. Assuming an initial state that is separable between the system and environment ($\mathbf{\rho}(0) = \mathbf{\rho}_{\rm S}(0) \otimes \mathbf{\rho}_{\rm E}(0) $), we can derive the system's non-unitary reduced dynamics by taking a partial trace over the environmental degrees of freedom:
\begin{align}
    \mathbf{\rho}_{\rm S}(t) & = \mathbf{Tr}_{\rm E} \left(\mathbf{U}(t,0) \rho(0) \mathbf{U}^\dagger(t,0) \right) \nl &= -\mathrm{i}[\mathbf{H}_{\rm S}, \mathbf{\rho}_{\rm S}(t)] + \mathcal{R} [\mathbf{\rho}_{\rm S}(t)],
\end{align}
where \red{$\mathbf{\rho}_{\rm S}(t)$} is the RDM of the system at time $t $, $\mathbf{Tr}_{\rm E} $ denotes the partial trace over the environment, \red{$\mathcal{R}$ is a superoperator that encodes the effects of the environment} and $\mathbf{U}(t,0) $ and $\mathbf{U}^\dagger(t,0) $ are the forward and backward time-evolution operators, respectively. Though most physical systems interact with some environment, exact solutions often become infeasible due to the exponential growth in complexity—a phenomenon known as the “curse of dimensionality”. Below, we introduce two commonly studied open quantum systems: the two-state spin-boson (SB) model and the Fenna-Matthews-Olson (FMO) complex.

\hspace{0.5pt}

\noindent\textit{SB model}: The SB model describes a qubit, or two-state system, coupled to a surrounding bath of independent harmonic oscillators. The Hamiltonian of the system, expressed in the basis of the excited ($\ket{e}$) and ground ($\ket{g}$) states, is:
\begin{equation}
    H = \epsilon \mathbf{\sigma}_z + \Delta \mathbf{\sigma}_x + \sum_{k} \omega_k \mathbf{b}_k^\dagger \mathbf{b}_k + \mathbf{\sigma}_z \sum_{k} c_k (\mathbf{b}_k^\dagger + \mathbf{b}_k),
\end{equation}
where $\mathbf{\sigma}_z$ and $\mathbf{\sigma}_x$ are Pauli matrices, $\epsilon$ is the energy difference between the two states, and $\Delta$ denotes their coupling strength. The environment consists of harmonic modes with creation and annihilation operators $\mathbf{b}_k^\dagger $ and $\mathbf{b}_k $ for each mode $k $, which has a frequency $\omega_k$. The interaction between the system and the environment is characterized by a coupling constant $c_k $ for each mode. The environmental influence is quantified by \red{Debye spectral density:}
\begin{equation} \label{eq:spectra}
    J(\omega) = 2 \lambda \frac{\gamma \omega}{\omega^2 + \gamma^2},
\end{equation}
where $\lambda$ is the reorganization energy and $\gamma$ is the characteristic frequency or the inverse of the relaxation time ($\gamma = 1/\tau$).

\hspace{0.5pt}

\noindent\textit{FMO Complex}: The FMO complex is a trimeric protein in green sulfur bacteria, critical to energy transfer in photosynthesis. Each monomer in the FMO complex includes several chlorophyll sites, typically seven or eight, which mediate energy transport. The dynamics within each monomer can be described by the Frenkel exciton model Hamiltonian: \red{
\begin{align}
    \mathbf{H} &= \sum_{i=1}^{n} \ket{i} \epsilon_i \bra{i} + \sum_{i \neq j}^{n} \ket{i} J_{ij} \bra{j} \nl & + \sum_{i=1}^{j} \sum_{k=1} \left(\frac{1}{2} \mathbf{P}_{k, i}^{2} + \frac{1}{2} \omega_{k, i}^{2} \mathbf{Q}_{k, i}^{2}\right) \mathbf{I} \nl & - \sum_{i=1}^{n} \sum_{k=1} \ket{i} c_{k,i} \mathbf{Q}_{k, i} \bra{i} + \sum_{i=1}^{n} \ket{i} \lambda_{i} \bra{i},
\end{align}}
where $n$ is the number of chlorophyll sites, $\epsilon_i$ is the site energy, and $J_{ij}$ is the coupling between sites $i$ and $j$. Here, $\mathbf{P}_{k,i}$ and $\mathbf{Q}_{k,i}$ are the momentum and position of the $k$-th mode interacting with site $i$, with $\omega_{k,i}$ representing the frequency of each mode. The $n\times n$ identity matrix $\mathbf{I}$ ensures dimensional consistency in the model. $c_{k,i} $ is the coupling constant between the $k$-th environmental mode and site $i$, and $\lambda_i$ is the reorganization energy for site $i$.

For FMO complex, we use the \red{Debye spectral density} (Eq.~\eqref{eq:spectra}), assuming it is identical for all sites.

\subsection{ML-enhanced quantum dissipative dynamics}

\red{In the machine learning framework, modeling the time evolution of an open quantum system with an $n$-dimensional Hilbert space can be viewed as learning a mapping function. This function, denoted as $\mathcal{M}$, takes a set of input descriptors and maps them to either a single RDM or a sequence of RDMs. Mathematically, this can be expressed as: 
\begin{equation}
    \mathcal{M}: \{\mathbb{R}^{i \times j}\}^k \to \{\mathbb{R}^{n \times n}\}^l \, ,
\end{equation}
where $\{\mathbb{R}^{i \times j}\}^k$ represents a collection of $k$ input matrices, each of size $i \times j$, which encode relevant physical information such as past history, initial conditions and simulation parameters. The output, $\{\mathbb{R}^{n \times n}\}^l$, consists of $l$ predicted RDMs, each of size $n \times n$, corresponding to different time steps. ML-based approaches developed so far can be divided into two main categories: recursive and non-recursive approaches, distinguished by the input descriptors and dynamics prediction.}

\hspace{0.5pt}

\noindent\textit{Recursive approaches}: \red{Recursive ML strategies\cite{ullah2021speeding, rodriguez2022comparative, herrera2021convolutional, herrera2024short} predict future dynamics based on their historical data, resembling traditional quantum dynamics where the evolution at any time step depends explicitly on the current state and implicitly on prior states. The recursive mapping function, denoted as $\mathcal{M}_{\rm rec}$, is defined as:  
\begin{equation}
     \mathcal{M}_{\rm rec}: \{\mathbb{R}^{n \times n}\}^k \to \mathbb{R}^{n \times n}
\end{equation}
such that  
\begin{equation}
\mathcal{M}_{\rm rec} \left[\mathbf{\rho}_{\rm S}(t_{k-k'}), \mathbf{\rho}_{\rm S}(t_{k-k'+1}), \dots, \mathbf{\rho}_{\rm S}(t_k)\right] = \mathbf{\rho}_{\rm S}(t_{k+1}),
\end{equation}
where $\mathcal{M}_{\rm rec}$ takes a sequence of $k^\prime$ previous RDMs, $\{\mathbf{\rho}_{\rm S}(t_{j})\}_{j=k-k^\prime}^k$, representing a short trajectory of the system’s past evolution. The function then predicts the RDM at the next time step, $\mathbf{\rho}_{\rm S}(t_{k+1}) \in \mathbb{R}^{n \times n}$. This process is applied iteratively: at each step, the newly predicted RDM is appended to the sequence, while the oldest entry is discarded, maintaining a fixed memory size of $k^\prime$ past time steps.}

\hspace{0.5pt}

\noindent\textit{Non-recursive approaches}:
Non-recursive approaches\cite{ullah2022predicting, ullah2022one} establish the mapping function $\mathcal{M}$ as a direct relationship between simulation parameters (e.g., time, temperature, coupling strength, initial conditions) and system dynamics, bypassing reliance on prior states. Non-recursive approaches can be further divided into two categories: time-dependent non-recursive approaches and time-independent non-recursive approaches. The former approaches employ a mapping function, $\mathcal{M}_{\rm AIQD}$, that depends explicitly on time $t$ and simulation parameters, defined as:
\begin{equation}
    \mathcal{M}_{\rm AIQD}: \mathbb{R} \times \mathbb{R}^p \to \mathbb{R}^{n \times n},
\end{equation}
such that
\begin{equation}
    \mathcal{M}_{\rm AIQD}(t, \mathbf{p}) = \mathbf{\rho}_{\rm S}(t),
\end{equation}
where $\mathbf{p}$ is a vector of simulation parameters. This methodology facilitates parallel computation of RDMs across all time steps, as each prediction is independent of prior results. The later time-independent non-recursive approaches,\cite{ullah2022one} on the other hand, determine the full RDM trajectory over a sequence of time steps in a single inference step. This is expressed as:
\begin{equation}
    \mathcal{M}_{\rm OSTL}: \mathbb{R}^p \to \{\mathbb{R}^{n \times n}\}^k, 
\end{equation}
such that
\begin{equation}
    \mathcal{M}_{\rm OSTL}(\mathbf{p}) = [\mathbf{\rho}_{\rm S}(t_1), \mathbf{\rho}_{\rm S}(t_2), \dots, \mathbf{\rho}_{\rm S}(t_k)],
\end{equation}
where $\mathbf{p}$ contains only the simulation parameters, and the model predicts the RDM trajectory for all specified time steps, $t_1, t_2, \dots, t_k$, in one computation. This approach streamlines trajectory prediction by eliminating the dependence on sequential steps.

\subsection{Expansion of RDM using traceless operators}
The RDM $\mathbf{\rho}_{\rm S}$ for a two-level quantum system can be elegantly expressed within the framework of $\mathfrak{su}(2)$, the Lie group of special unitary transformations in two dimensions. In this context, $\mathbf{\rho}_{\rm S}$ is decomposed as a linear combination of the identity operator and the generators of the $\mathfrak{su}(2)$ Lie algebra, the Pauli matrices:

\begin{equation}
\mathbf{\rho}_{\rm S} = a_0 \mathbf{I} + a_x \mathbf{\sigma}_x + a_y \mathbf{\sigma}_y + a_z \mathbf{\sigma}_z = a_0 \mathbf{I} + \mathbf{a} \cdot \mathbf{\sigma},
\end{equation}
where $a_0, a_x, a_y,$ and $a_z$ are real coefficients, and $\mathbf{a} = (a_x, a_y, a_z)$ serves as a time-dependent vector of coefficients. 
\red{The basis elements (generators) $\mathbf{\sigma} = (\mathbf{\sigma}_x, \mathbf{\sigma}_y, \mathbf{\sigma}_z)$ 
are traceless Hermitian matrices:}

\begin{equation}
\mathbf{\sigma}_x = \frac{1}{2}\begin{pmatrix} 0 & 1 \\ 1 & 0 \end{pmatrix}, \quad 
\mathbf{\sigma}_y = \frac{1}{2}\begin{pmatrix} 0 & -\mathrm{i} \\ \mathrm{i} & 0 \end{pmatrix}, \quad 
\mathbf{\sigma}_z = \frac{1}{2}\begin{pmatrix} 1 & 0 \\ 0 & -1 \end{pmatrix}.
\end{equation}

The decomposition above reflects the structure of $\mathfrak{su}(2)$, with $\mathbf{I}$ as the trivial (identity) representation. Three key mathematical properties of the $\mathfrak{su}(2)$ algebra are fundamental to this formulation: (1) The generators of $\mathfrak{su}(2)$, $\mathbf{\sigma}_i$, are traceless, i.e., $\mathbf{Tr}[\mathbf{\sigma}_i] = 0$. In contrast, the identity matrix $\mathbf{I}$ has $\mathbf{Tr}[\mathbf{I}] = 2$. This tracelessness ensures that the coefficient \red{$a_0=1/n$} determines the trace of $\mathbf{\rho}_{\rm S}$, while the remaining coefficients, $a_x, a_y, a_z$ correspond to the traceless part of $\mathbf{\rho}_{\rm S}$. This separation into trace and traceless components arises naturally from the $\mathfrak{su}(2)$ structure; (2) The Pauli matrices are orthogonal under the trace inner product, i.e., $\mathbf{Tr}[\mathbf{\sigma}_i \mathbf{\sigma}_j] = \frac{1}{2} \delta_{ij}$. This orthogonality implies that the coefficients $a_x, a_y, a_z$ are uniquely determined by projecting $\mathbf{\rho}_{\rm S}$ onto the basis elements of $\mathfrak{su}(2)$, i.e., $a_i=2\mathbf{Tr}[\mathbf{\rho}_{\rm S}\sigma_i]$; (3) The quadratic Casimir operator, which is central to the representation theory of $\mathfrak{su}(2)$, takes the form:
\begin{equation}
\sum_{i=1}^3 \mathbf{\sigma}_i^2 = \frac{3}{4} \mathbf{I}.
\end{equation}
This result reflects the fact that the sum of the squares of the generators is proportional to the identity operator. 

Building upon the formalism developed for two-level systems ($n=2$), we now generalize the representation of the RDM and associated operators to $n$-level quantum systems. This extension utilizes the mathematical framework of the $\mathfrak{su}(n)$ Lie algebra. In such systems, any Hermitian $n \times n$ matrix can be expressed as a linear combination of the identity matrix $\mathbf{I}$ and the traceless generators of $\mathfrak{su}(n)$. This decomposition is given by:  
\begin{equation}
\mathbf{\rho}_{\rm S} = a_0 \mathbf{I} + \sum_{i=1}^{n^2-1} a_i \mathfrak{S}_i,
\end{equation}  
where $\mathbf{I}$ is the $n \times n$ identity matrix, $\mathfrak{S}_i$ ($i = 1, \dots, n^2-1$) are traceless Hermitian matrices that serve as the generators of $\mathfrak{su}(n)$, and \red{$a_i$} are coefficients encoding the system dynamics. These generators satisfy the same orthogonality relations as those in the $\mathfrak{su}(2)$ algebra, $\mathrm{tr}[\mathfrak{S}_i\mathfrak{S}_j]=\frac{1}{2}\delta_{ij}$.  

Furthermore, the quadratic Casimir operator maintains the same proportionality relationship with the identity operator as in the $\mathfrak{su}(2)$ case:  
\begin{equation}
\sum_{i=1}^{n^2-1} \mathfrak{S}_i^2 = \frac{n^2-1}{2n} \mathbf{I}.
\end{equation}  
This operator is invariant under unitary transformations and independent of the specific choice of $\mathfrak{su}(n)$ basis, highlighting the universal properties of the Lie algebra representation.

The $\mathfrak{su}(n)$ basis (generators) can be constructed as a generalization of the Pauli matrices from $\mathfrak{su}(2)$ and is systematically divided into three categories: $n(n-1)/2$ symmetric off-diagonal matrices, $n(n-1)/2$ antisymmetric off-diagonal matrices, and $n-1$ diagonal matrices. The symmetric off-diagonal basis are defined as:   
\begin{equation}
\mathfrak{S}_{ij}^+ = \frac{1}{2} \left(\ket{i} \bra{j} + \ket{j} \bra{i}\right), \quad 1 \leq i < j \leq n,
\end{equation}  
while the antisymmetric off-diagonal basis are given by:  
\begin{equation}
\mathfrak{S}_{ij}^- = -\frac{\mathrm{i}}{2} \left(\ket{i} \bra{j} - \ket{j} \bra{i}\right), \quad 1 \leq i < j \leq n.
\end{equation}  
The diagonal basis are constructed as:  
\begin{equation}
\mathfrak{S}_j = \sqrt{\frac{1}{2j(j-1)}} \left( \sum_{k=1}^{j-1} \ket{k}\bra{k} + (1-j) \ket{j}\bra{j} \right), \quad 2 \leq j \leq n.
\end{equation}  

It is important to emphasize that this choice for the $\mathfrak{su}(n)$ basis is not unique. Any traceless Hermitian basis that satisfies the commutation and normalization conditions can be employed.

\section{Loss function, data preparation and training}

 To showcase the effectiveness of our $\mathfrak{su}(n)$-bases approaches, we examine the relaxation dynamics of the SB model and exciton energy transfer (EET) in the FMO complex. For comparison, we utilize four models: Purely Data-driven Physics-Uninformed Neural Networks (PUNN), $\mathfrak{su}(n)$-based PUNN ($\mathfrak{su}(n)$-PUNN), Physics-Informed Neural Networks (PINN), and $\mathfrak{su}(n)$-based PINN ($\mathfrak{su}(n)$-PINN). All models employ a hybrid architecture comprising convolutional neural networks (CNN), followed by long short-term memory (LSTM) layers, and fully connected dense layers (CNN-LSTM). Optimization was carried out using a composite loss function given, in general, by  

\begin{equation}
    \mathcal{L} = \alpha_1\mathcal{L}_1 + \alpha_2\mathcal{L}_2 + \alpha_3\mathcal{L}_3 + \alpha_4\mathcal{L}_4 + \alpha_5\mathcal{L}_5\, ,
\end{equation}  
where each term is defined as follows. The term $\mathcal{L}_1$ represents the mean squared error (MSE) between the predicted ($\mathbf{\rho}_{\text{S}}$) and reference ($\tilde{\mathbf{\rho}}_{\text{S}}$) RDM's elements,  

\begin{equation}
  \mathcal{L}_1 = \frac{1}{N_t \cdot n^2} \sum_{t=1}^{N_t} \sum_{i, j = 1}^{n} \left(\tilde{\mathbf{\rho}}_{\text{S}, ij}(t) - \mathbf{\rho}_{\text{S}, ij}(t) \right)^2 \, ,
\end{equation}  
where $N_t$ is the number of time steps. The term $\mathcal{L}_2$ enforces the trace constraint by penalizing deviations of the trace of the predicted RDMs from unity,  

\begin{equation}
  \mathcal{L}_2 = \frac{1}{N_t} \sum_{t=1}^{N_t} \left( \mathbf{Tr} \, \mathbf{\rho}_{\text{S}}(t) - 1 \right)^2 \, .  
\end{equation}  

To ensure that the predicted RDMs are Hermitian, the deviation from Hermitian symmetry is penalized as:
\begin{equation}
  \mathcal{L}_3 = \frac{1}{N_t \cdot n^2} \sum_{t=1}^{N_t} \sum_{i, j = 1}^{n} \abs{\mathbf{\rho}_{\text{S}, ij}(t) - 
  \mathbf{\rho}_{\text{S}, ji}(t)^*}^2 \, .
\end{equation}  

The term $\mathcal{L}_4$ ensures positive semi-definiteness of the density matrix by penalizing negative eigenvalues $\mu_i(t)$,  

\begin{equation}
  \mathcal{L}_4 = \frac{1}{N_t \cdot n} \sum_{t=1}^{N_t} \sum_{i=1}^{n} \mathbf{\text{max}}(0, -\mu_i(t))^2 \, ,  
\end{equation}  
where, for a negative eigenvalue, the term $-\mu_i(t)$ becomes positive, resulting in $\mathbf{\text{max}}(0, -\mu_i(t)) = -\mu_i(t)$, while it evaluates to 0 otherwise. The subsequent loss term, $\mathcal{L}_5$ constrains the eigenvalues to remain within the range $[0, 1]$, \red{a fundamental property of all realistic RDMs}.  
\red{
\begin{equation}
  \mathcal{L}_5 = \frac{1}{N_t \cdot n} \sum_{t=1}^{N_t} \sum_{i=1}^{n} \left(\mathbf{\text{clip}}\left(\mu_i(t), 0, 1\right) - \mu_i(t)\right)^2 \, ,
\end{equation}  }
where the clip function clips $\mu_i(t)$ to the interval $[0, 1]$
\begin{equation}
\text{clip}(\mu_i(t), 0, 1) = 
\begin{cases} 
0, & \text{if } \mu_i(t) < 0, \\
\mu_i(t), & \text{if } 0 \leq \mu_i(t) \leq 1, \\
1, & \text{if } \mu_i(t) > 1.
\end{cases}
\end{equation}

\begin{figure}[!thb]
    \centering
    \includegraphics[width=0.5\textwidth]{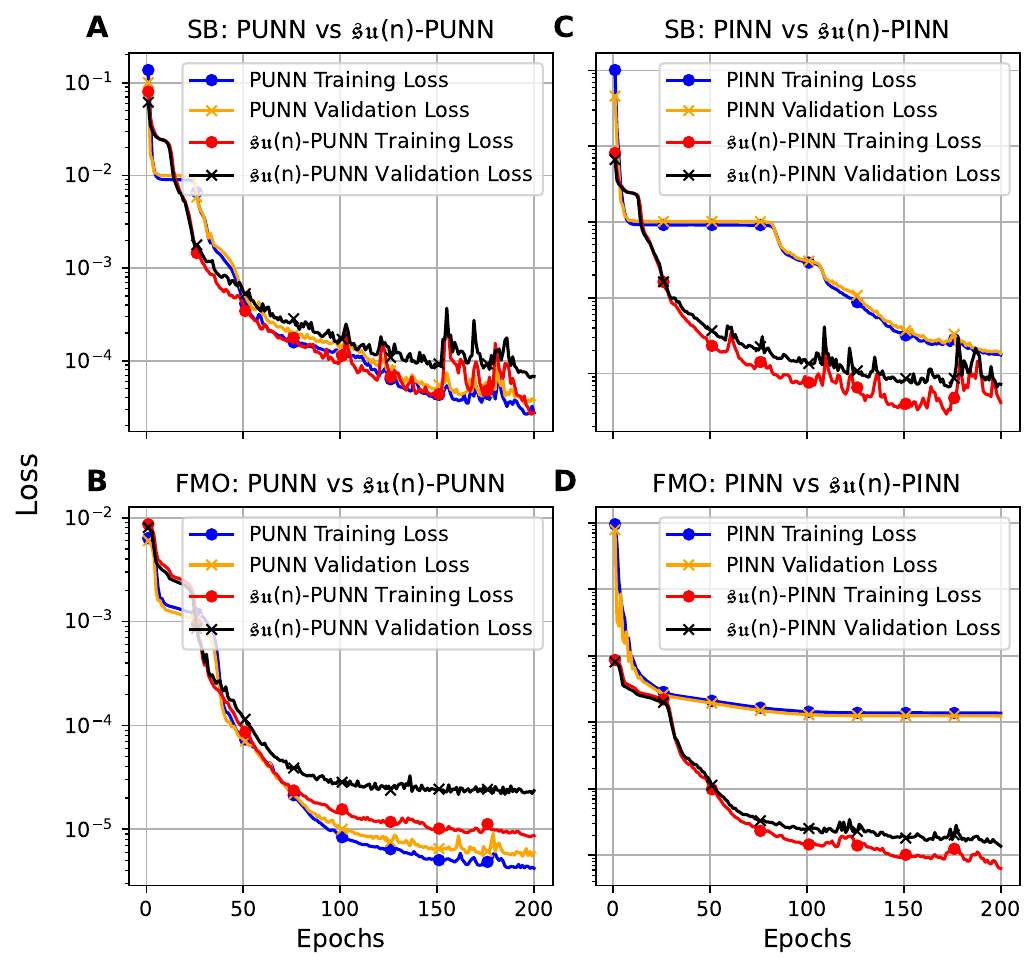}
    \caption{Comparison of training and validation losses for PUNN, \(\mathfrak{su}(n)\)-PUNN, PINN, and \(\mathfrak{su}(n)\)-PINN as a function of the number of epochs. Panels A and B compare PUNN with \(\mathfrak{su}(n)\)-PUNN for the SB (Panel A) and FMO (Panel B) models, respectively. Panels C and D compare PINN with \(\mathfrak{su}(n)\)-PINN for the SB (Panel C) and FMO (Panel D) models, respectively.}
    \label{fig:lr_curves}
\end{figure}

The coefficients $\alpha_1, \alpha_2, \alpha_3, \alpha_4, \alpha_5$ are tuning parameters that determine the relative contributions of the respective loss terms. Together, the terms $\mathcal{L}_1$, $\mathcal{L}_2$, $\mathcal{L}_3$, $\mathcal{L}_4$ and $\mathcal{L}_5$ ensure the validity of the RDM by enforcing accuracy, trace conservation, hermiticity, positive semi-definiteness, and eigenvalue constraints, respectively. For both the PUNN and $\mathfrak{su}(n)$-PUNN models, only $\mathcal{L}_1$ is included with $\alpha_1 = 1$ and $\alpha_2 = \alpha_3 = \alpha_4 = \alpha_5 = 0$, meaning no explicit physical constraints are applied. In the case of PINN, all constraints are employed with $\alpha_1 = \alpha_2  = 1.0$, $\alpha_3 = 2.0$ and $\alpha_4 = \alpha_5  = 3.0$. Similarly, the $\mathfrak{su}(n)$-PINN model incorporates $\mathcal{L}_1$, $\mathcal{L}_3$, $\mathcal{L}_4$ and $\mathcal{L}_5$ with the same tuned values as for PINN. Trace conservation is inherently satisfied in this case due to the properties of the $\mathfrak{su}(n)$ Lie algebra representation, eliminating the need for $\mathcal{L}_2$ (i.e., $\alpha_2 = 0$).  

To train our models, for the SB model, the training data were sourced from the publicly accessible QD3SET-1 database,\cite{ullah2023qd3set} which provides precomputed RDMs via the hierarchical equations of motion (HEOM) approach.\cite{tanimura1989time, shi2009efficient, chen2022universal, xu2022taming} The dataset, denoted as $\mathcal{D}_{\mathrm{sb}}$, includes 1000 simulations 
over a four-dimensional parameter space ($\varepsilon / \Delta, \lambda / \Delta, \gamma / \Delta$, and $\beta \Delta$) that defines the system-bath coupling strength, bath reorganization energy, bath relaxation rate, and inverse temperature. Similarly, for the seven-site FMO complex, data from the QD3SET-1 database were used. This dataset consists of 1000 simulations 
describing the dynamics for initial excitations at site-1 and site-6, spanning a parameter space $(\lambda, \gamma, T)$. Dynamics were propagated using the trace-conserving local thermalizing Lindblad master equation (LTLME),\cite{mohseni2008environment} with the Hamiltonian parameterized based on the work of Adolphs and Renger.\cite{adolphs2006proteins}  


To train the models, we employed OSTL-based dynamics propagation where RDM $\rho_{\rm {S}}(t)$ and coefficients $a_i$ at each time step was converted into a 1D vector. In both PUNN and PINN models, the Hermitian property $\mathbf{\rho}_{\rm {S},{ij}}(t) = \mathbf{\rho}_{\rm {S},{ji}}(t)^*$ is explicitly utilized. Consequently, only the diagonal elements (considering their real parts) and the upper off-diagonal elements were included in the analysis, with the real and imaginary components of the off-diagonal terms treated separately. As a result, the Hermiticity loss term $\mathcal{L}_3$ is zero by definition in these models and does not affect training. While not strictly necessary in these cases, we retained it to maintain consistency across model types and for pedagogical clarity. Further methodological details can be found in the Supporting Information. To optimize training process, farthest point sampling\cite{dral2019mlatom, ullah2022predicting} was applied to select a subset of trajectories. For each case of the SB model ($\varepsilon / \Delta =0$ and $1$) and the FMO complex (site-1 and 6), 400 trajectories were selected for training, and the remaining data were reserved for testing. Separate CNN-LSTM models were trained for the SB model and the FMO complex, each using an identical architecture. For comparison, we selected models with comparable training and validation losses.

\begin{figure}[!thb]
    \centering
    \includegraphics[width=0.5\textwidth]{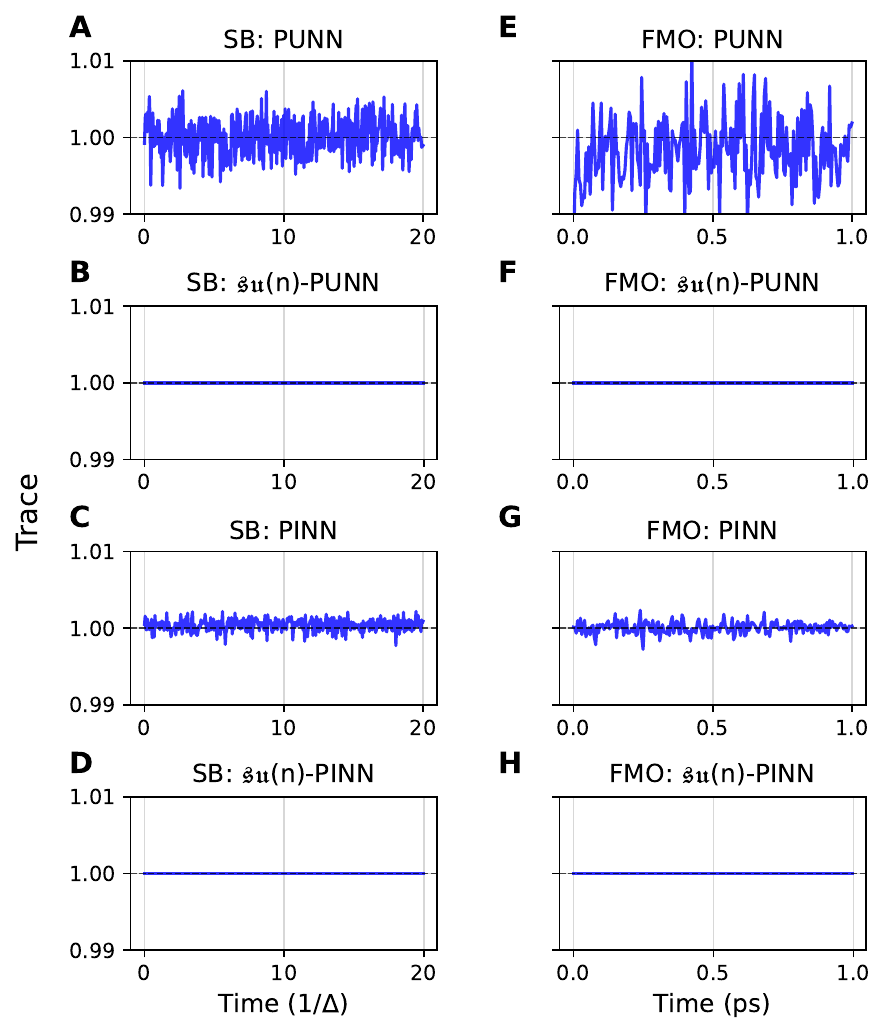}
    \caption{Comparison of trace conservation in quantum dissipative dynamics using PUNN (A, E), $\mathfrak{su}(n)$-PUNN (B, F), PINN (C, G), and $\mathfrak{su}(n)$-PINN (D, H). The first column (A-D) presents results for the asymmetric SB model, while the second column (E-H) focuses on the FMO complex with initial excitation on site 1. For the asymmetric SB model, the results correspond to an unseen dynamics characterized by $\varepsilon / \Delta = 1.0$, $\gamma/\Delta = 9.0$, $\lambda/\Delta = 0.6$, and $\beta\Delta = 1.0$. For the FMO complex, the parameters are $\gamma = 400~\text{cm}^{-1}$, $\lambda = 40~\text{cm}^{-1}$, and $T = 90~\text{K}$. Details on training and prediction are provided in the Results and Discussion section.}
    \label{fig:trace}
\end{figure}

\section{Results and Discussion}
In this section, we demonstrate the results including trace conservation, eigenvalue spectra, predicted dynamics and accuracy for all the models. 
This section presents a comprehensive evaluation of the proposed models. We analyze and compare the performance of PUNN, $\mathfrak{su}(n)$-PUNN, PINN, and $\mathfrak{su}(n)$-PINN models in terms of learning efficiency, trace conservation, positive semi-definiteness, and dynamical accuracy for both the SB model and the FMO complex. Additionally, we investigate the influence of NN architectures and explore an alternative variant of PUNN, termed TC-PUNN, which enforces trace conservation by design.

\subsection{Comparison of PUNN, $\mathfrak{su}(n)$-PUNN, PINN, and $\mathfrak{su}(n)$-PINN}
We begin by comparing the performance of the four primary model formulations: PUNN, $\mathfrak{su}(n)$-PUNN, PINN, and $\mathfrak{su}(n)$-PINN. Figure~\ref{fig:lr_curves} shows the learning curves, highlighting the decrease in training loss over epochs for each model. For the SB model, both PUNN and $\mathfrak{su}(n)$-PUNN exhibit comparable learning performance. However, $\mathfrak{su}(n)$-PINN demonstrates significantly superior learning efficiency across all epochs, with the loss function decreasing rapidly. In the case of the more complex FMO system, PUNN achieves a marginally lower loss compared to $\mathfrak{su}(n)$-PUNN. Nonetheless, similar to the SB model, $\mathfrak{su}(n)$-PINN outperforms PINN in terms of learning efficiency. These findings highlight that expanding the RDM in the $\mathfrak{su}(n)$ basis simplifies optimization and enhances learning efficiency and can describe more physically realistic behavior.

Fig.~\ref{fig:trace} highlights the performance of trace conservation across the four models. The purely data-driven PUNN model fails entirely to conserve the trace in both cases. While PINN shows substantial improvement in trace conservation, minor violations still occur. This limitation stems from the fact that the physical constraints in the PINN loss function are considered to be "soft" constraints. As a result, they guide the training process but are not strictly enforced.\cite{ullah2024pinn, wang2021physics, norambuena2024physics} In contrast, the $\mathfrak{su}(n)$-PUNN and $\mathfrak{su}(n)$-PINN ensure the exact trace conservation by embedding the trace explicitly via the identity matrix, as a result achieving perfect trace preservation by design, rather than relying on the loss function.

\begin{figure}[!thb]
    \centering
    \includegraphics[width=0.5\textwidth]{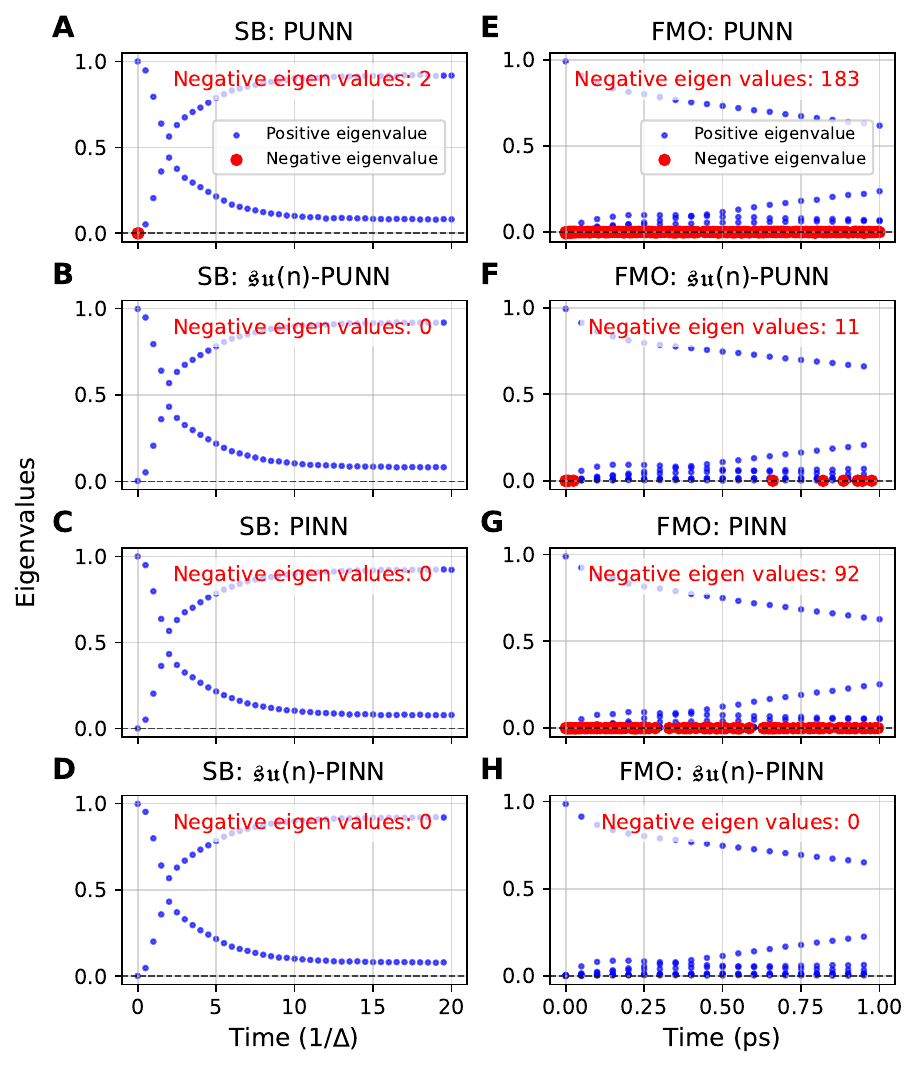}
    \caption{Evaluation of positive semi-definiteness and eigenvalue spectra of the predicted RDMs using PUNN (A, E), $\mathfrak{su}(n)$-PUNN (B, F), PINN (C, G), and $\mathfrak{su}(n)$-PINN (D, H) for the SB model (A–D) and the FMO complex (E–H). Sparse positive eigenvalues are shown for clarity, and the number of negative eigenvalues is indicated for each model. Simulation parameters are consistent with those in Fig.~\ref{fig:trace}.}
    \label{fig:eig}
\end{figure}

To evaluates the performance Fig.~\ref{fig:eig} evaluates the positive semi-definiteness and eigenvalue constraints of the predicted RDMs by analyzing their eigenvalues. For simpler systems, such as the SB model, all models perform similarly, with nearly all eigenvalues remaining positive, except for a few negative eigenvalues in the PUNN results. In more complex cases, like the FMO complex, the PUNN model produces approximately 13\% negative eigenvalues (183 out of 1407), while PINN yields 3.6\% negative eigenvalues (52 out of 1407), demonstrating that incorporating the loss terms $\mathcal{L}_4$ and $\mathcal{L}_5$ improves the positivity of the predicted RDMs. In contrast, the purely data-driven $\mathfrak{su}(n)$-PUNN results in only 0.78\% negative eigenvalues (11 out of 1407), outperforming PINN even with all constraints included. Finally, by adding the corresponding loss terms in $\mathfrak{su}(n)$-PINN, all predicted RDMs achieve complete positivity. This further confirms that our $\mathfrak{su}(n)$ approach enhances learning efficiency.

\begin{table}[ht]
\scriptsize
\centering
\caption{Average mean absolute error (MAE) for the diagonal (Diag) and off-diagonal (Off-diag) elements of the RDM predicted by the PUNN, $\mathfrak{su}(n)$-PUNN, PINN, and $\mathfrak{su}(n)$-PINN models for the SB model and FMO complex. The off-diagonal errors include the average MAE for both the real and imaginary components.}
\label{tab:mae_results}
\begin{tabular}{lcccc}
\hline
\textbf{Model} & \multicolumn{2}{c}{\textbf{SB Model}} & \multicolumn{2}{c}{\textbf{FMO Complex}} \\
\cline{2-3} \cline{4-5}
               & \textbf{Diag} & \textbf{Off-diag} & \textbf{Diag} & \textbf{Off-diag} \\ 
               &                    & \textbf{(Real, Imag)}  &                    & \textbf{(Real, Imag)} \\ 
\hline
PUNN           & 0.0032             & (0.0026, 0.0027)       & 0.0074             & (0.0031, 0.0014)       \\ 
$\mathfrak{su}(n)$-PUNN & 0.0020     & (0.0011, 0.0013)       & 0.0056             & (0.0019, 0.00077)       \\ 
PINN           & 0.0027             & (0.0028, 0.0018)       & 0.0092             & (0.0021, 0.00072)       \\ 
$\mathfrak{su}(n)$-PINN & 0.0017     & (0.0014, 0.0014)       & 0.0037             & (0.0014, 0.00071)       \\ 
\hline
\end{tabular}
\end{table}

To further assess the performance of the four models, Table~\ref{tab:mae_results} compares the accuracy of PUNN, $\mathfrak{su}(n)$-PUNN, PINN, and $\mathfrak{su}(n)$-PINN in predicting the evolution of RDM elements (population and coherence terms) for the SB model and the FMO complex, as illustrated in Fig.~S1 (Supporting Information). For the SB model, all models provide accurate predictions; however, $\mathfrak{su}(n)$-PUNN and $\mathfrak{su}(n)$-PINN achieve the lowest errors for both diagonal and off-diagonal elements, demonstrating superior learning capabilities compared to PUNN and PINN. Although PUNN and PINN also perform well, their errors are marginally higher.

In the case of the FMO complex, $\mathfrak{su}(n)$-PINN exhibits the best overall performance, followed by $\mathfrak{su}(n)$-PUNN. Similar to the SB model, PUNN and PINN show higher errors for both diagonal and off-diagonal terms.

\begin{figure}[!thb]
    \centering
    \includegraphics[width=0.5\textwidth]{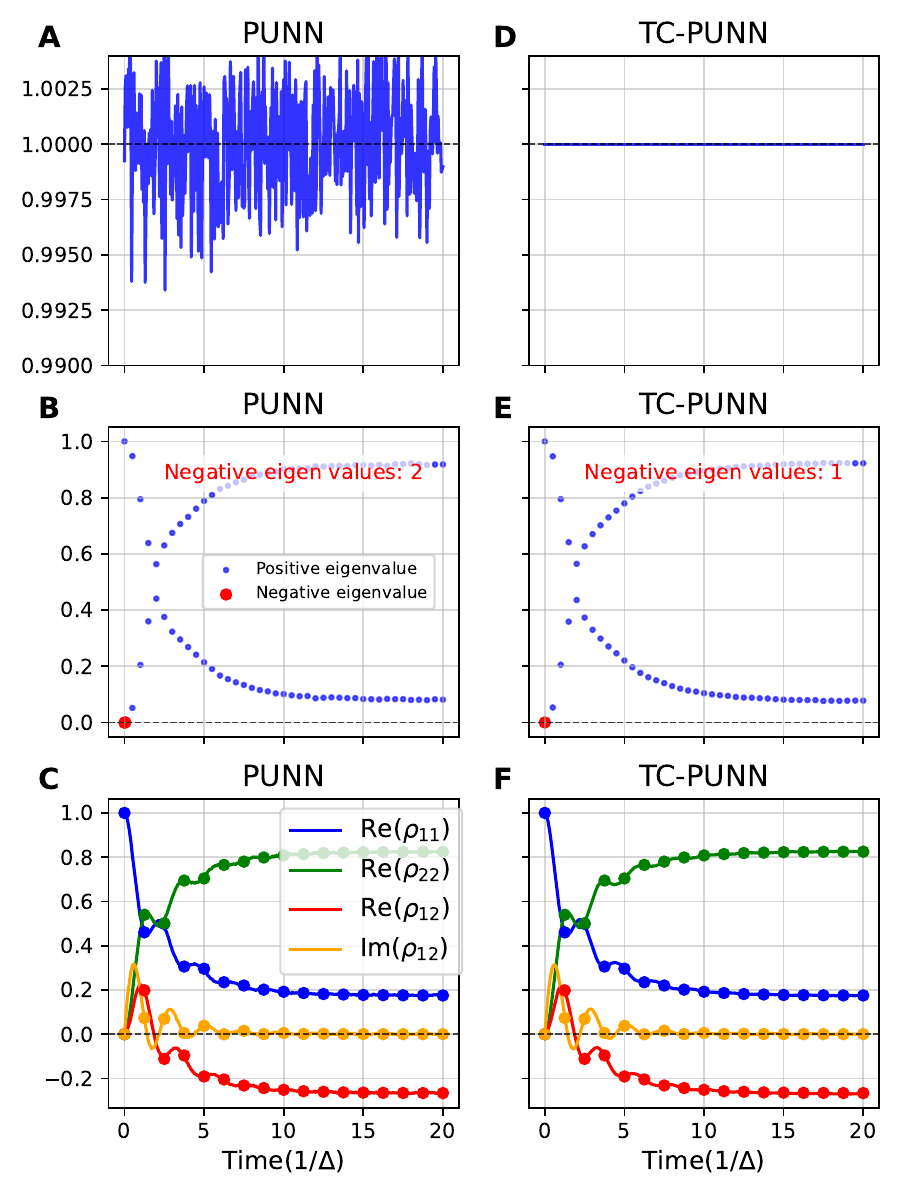}
    \caption{Comparison of PUNN (A–C) and TC-PUNN (D–F) models for the asymmetric SB model. Panel A and D show trace conservation; panels B and E show eigenvalue spectra of the predicted RDMs; and panels C and F depict the time evolution of representative RDM elements, including both populations and coherences. Reference dynamics is overlaid as dots. The simulation parameters are given in the caption of Fig.~\ref{fig:trace}.}
    \label{fig:tc_punn_sb_trace}
\end{figure}

\begin{figure}[!thb]
    \centering
    \includegraphics[width=0.5\textwidth]{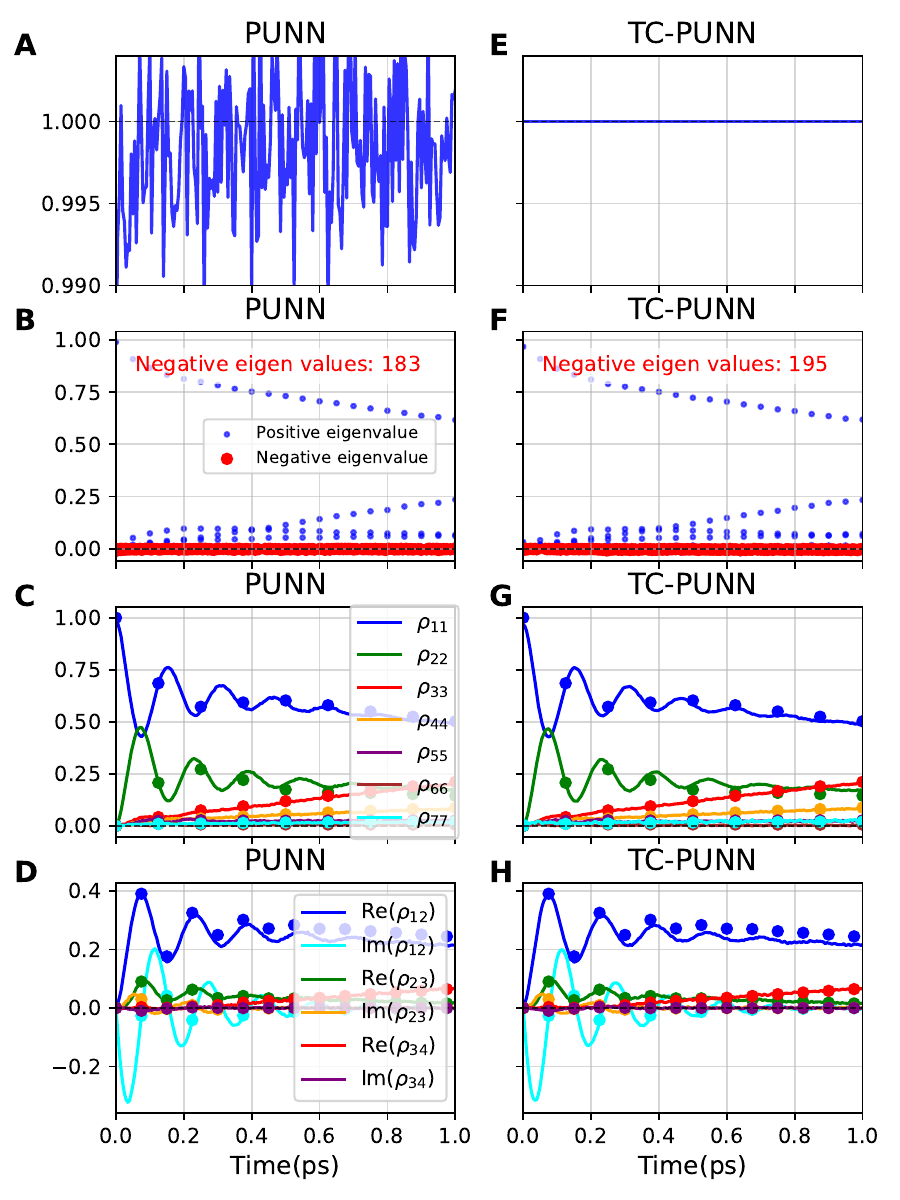}
    \caption{Comparison of PUNN (panels A–D) and TC‑PUNN (panels E–H) models for the FMO complex with initial excitation on site-1. Panels A and E show trace conservation; panels B and F show eigenvalue spectra of the predicted RDMs; panels C and G depict the time evolution of diagonal RDM elements (populations); and panels D and H depict the time evolution of selected off‑diagonal RDM elements (coherences). Reference dynamics is overlaid as dots. The simulation parameters are given in the caption of Fig.~\ref{fig:trace}.}
    \label{fig:tc_punn_fmo_trace}
\end{figure}

\subsection{PUNN and PINN with explicit trace conservation}

As with the $\mathfrak{su}(n)$-based models, exact trace conservation can also be enforced in traditional PUNN and PINN architectures by construction. It can be achieved by excluding one of the diagonal elements (typically the $n$th) during training, and reconstructing it using the constraint $\rho_{nn,\text{S}} = 1 - \sum_{i=1}^{n-1} \rho_{ii,\text{S}}$. This approach guarantees that the trace of the RDM remains exactly unity at all time steps. We demonstrate this strategy only for PUNN—referred to as PUNN with exact trace conservation (TC-PUNN)—in both the SB model and FMO complex, with results shown in Fig.~\ref{fig:tc_punn_sb_trace} and Fig.~\ref{fig:tc_punn_fmo_trace}, respectively.

To evaluate the performance of TC-PUNN, we compare it against the standard PUNN model. While TC-PUNN ensures strict trace preservation by construction, it does not necessarily lead to improved overall predictive accuracy—especially for larger systems like the FMO complex, as summarized in Table~\ref{tab:tc_punn_mae}. In the SB model, which involves only a single diagonal element, the use of exact trace conservation yields a notable improvement in accuracy for the diagonal elements. However, in larger systems such as the FMO complex, omitting a diagonal element from training and inferring it via the trace condition does not significantly improve performance. The benefit of exact trace conservation is diminished as the complexity of the system increases.

When comparing the eigenvalues spectra of the predicted RDMs, TC-PUNN shows a reduction in the number of negative eigenvalues for the SB model, indicating better physical consistency. However, this advantage does not extend to the FMO complex; in fact, the number of negative eigenvalues increases in this case (see Fig.~\ref{fig:tc_punn_fmo_trace}F), suggesting that enforcing exact trace conservation may come at the cost of overall matrix consistency.

It is also important to note that while the final diagonal element is constrained to enforce trace conservation, its predicted value can diverge significantly from the true value, especially when the errors in the remaining $n-1$ diagonal elements accumulate. Thus, although TC-PUNN guarantees trace conservation, it does not always lead to better accuracy or physical fidelity across all systems.

\begin{table}[h]
\scriptsize
\centering
\caption{Comparison of PUNN and TC-PUNN models based on the average MAE for the diagonal (Diag) and off-diagonal (Off-diag) elements of the RDM predicted for the SB model and FMO complex. The off-diagonal errors are computed as the average MAE for both the real and imaginary components of the RDM elements.}
\label{tab:tc_punn_mae}
\begin{tabular}{l c c c c}
\toprule
\textbf{Model} & \multicolumn{2}{c}{\textbf{SB Model}} & \multicolumn{2}{c}{\textbf{FMO Complex}} \\
\cline{2-3} \cline{4-5}
               & \textbf{Diag} & \textbf{Off-diag} & \textbf{Diag} & \textbf{Off-diag} \\ 
               &                    & \textbf{(Real, Imag)}  &                    & \textbf{(Real, Imag)} \\ 
\midrule
PUNN               & 0.0032     & (0.0026,  0.0027)    &  0.0074     & (0.0031,  0.0014)  \\ 
TC-PUNN              & 0.0019   & (0.0026,  0.0017)    &  0.0080     & (0.0031,  0.0013)   \\
\bottomrule
\end{tabular}
\end{table}

\begin{figure}[!thb]
    \centering
    \includegraphics[width=0.5\textwidth]{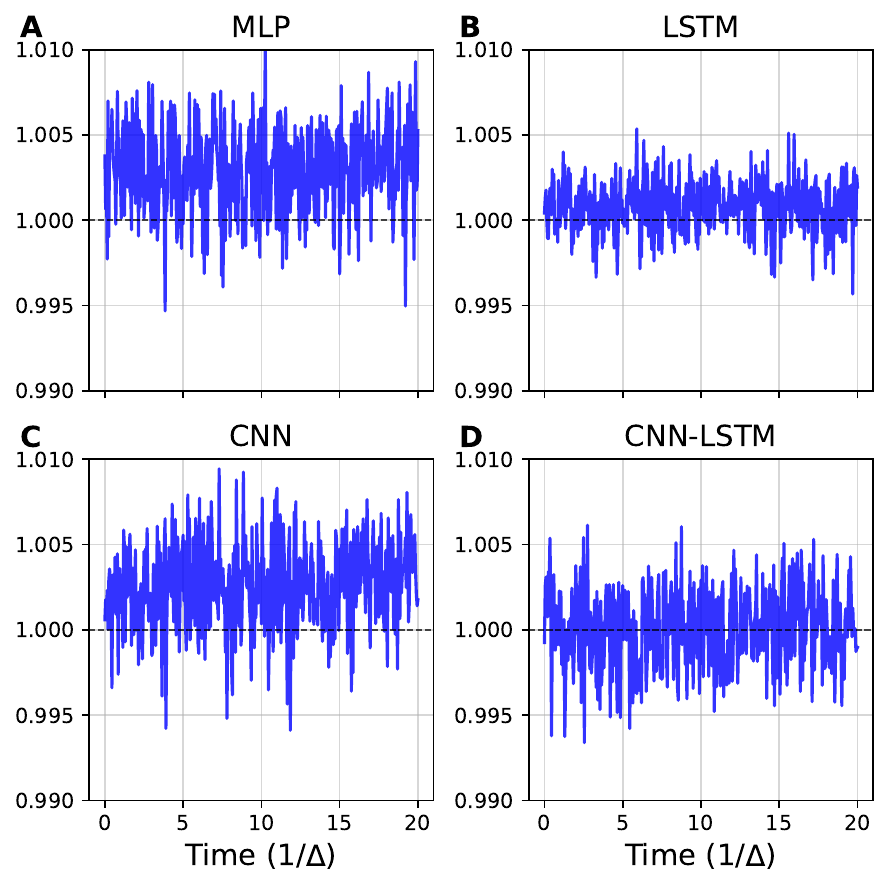}
    \caption{Comparison of trace conservation in quantum dissipative dynamics for the asymmetric SB model using the PUNN approach with different neural network architectures: (A) MLP, (B) LSTM, (C) CNN, and (D) CNN-LSTM. All models are evaluated on an unseen test case with parameters given in the caption of Fig.~\ref{fig:trace}.}
    \label{fig:compare_nn_sb_trace}
\end{figure}

\begin{table}[h]
\centering
\caption{Comparison of different NN model architectures based on the average MAE for the diagonal (Diag) and off-diagonal (Off-diag) elements of the RDM predicted by the PUNN approach using MLP, LSTM, CNN, and CNN-LSTM models for the SB model. The off-diagonal errors are calculated as the average MAE for both the real and imaginary components of the RDM elements.}
\label{tab:compare_nn_mae_results}
\begin{tabular}{l c c}
\toprule
\textbf{Model} & \textbf{Diag} & \textbf{Off-diag} \\

               &                & \textbf{(Real, Imag)}  \\ 
\midrule
MLP               & 0.0030     & (0.0019,  0.0021)     \\ 
LSTM              & 0.0019     & (0.0016,  0.0016)     \\ 
CNN               & 0.0032     & (0.0035, 0.0028)      \\ 
CNN-LSTM          & 0.0032     & (0.0026, 0.0027)      \\ 
\bottomrule
\end{tabular}
\end{table}

\subsection{Impact of NN architecture: MLP, LSTM, CNN, and CNN-LSTM}

To investigate the influence of NN architecture on the modeling of quantum dissipative dynamics, we compare four common architectures—Multilayer perceptron (MLP), LSTM, CNN, and a hybrid CNN-LSTM—within the PUNN framework. This comparison is conducted using the SB model, evaluating each architecture in terms of trace conservation, dynamical accuracy, and positive semi-definiteness.

As shown in Fig.~\ref{fig:compare_nn_sb_trace}, the LSTM architecture, which is well-suited for temporal sequence modeling, demonstrates superior performance in preserving the trace, with comparative small fluctuations throughout the dynamics. CNN-LSTM follows closely, while CNN and MLP show larger deviations, with MLP performing the worst in this regard.

For the positive semi-definiteness of the predicted RDMs (Fig.~S2, Supporting Information), LSTM and CNN-LSTM again outperform the other architectures, followed by CNN and MLP. When assessing the predictive accuracy of RDM dynamics, depicted in Fig.~S3 (Supporting Information) and quantified by MAE in Table~\ref{tab:compare_nn_mae_results}, the LSTM model achieves the lowest errors. Interestingly, MLP performs better in dynamical accuracy than CNN-LSTM and CNN, despite being inferior in enforcing physical constraints.

These results suggest that while architectures like LSTM are more effective in capturing both physical constraints and accurate dynamics, there can be a disconnect between physical fidelity (e.g., trace conservation, positivity) and pure prediction accuracy—highlighting a trade-off that must be considered when selecting an appropriate neural network structure for quantum dynamics modeling.

\section{Final remarks and future directions}
In this work, we addressed key limitations of conventional ML approaches for modeling RDMs, particularly their challenges in training efficiency and the enforcement of physical constraints such as trace conservation. Traditional methods often rely on directly modeling the full RDM and impose physical constraints via complex nonlinear loss terms, which can complicate optimization and lead to unphysical predictions.

To overcome these issues, we proposed a novel framework based on the $\mathfrak{su}(n)$ Lie algebra, which expresses the RDM as a linear combination of the identity operator and $n^2 - 1$ orthonormal, Hermitian, and traceless generators. The traceless property of the generators guarantees that only the identity component contributes to the trace, thereby ensuring exact trace conservation by construction. Unlike PINN or traditional models that require explicit penalty terms in the loss function to enforce constraints, our approach simplifies optimization and improves training stability. Furthermore, the algebraic formulation offers a natural path for generalizing ML models to quantum systems of varying dimensionality.

In addition to the core models presented in the main text, we also explored an alternative formulation, TC-PUNN, which enforces exact trace conservation in a conventional PUNN setting by reconstructing the final diagonal element of the RDM. While this variant succeeds in guaranteeing trace preservation, it does not consistently improve predictive accuracy or physical consistency across all systems—particularly in higher-dimensional settings, where error propagation can degrade the overall fidelity of the predictions.

Moreover, we conducted a comparative study of different neural network architectures—MLP, LSTM, CNN, and CNN-LSTM—within the PUNN framework to assess the impact of network structure on performance. The results reveal that architectures with temporal modeling capabilities, particularly LSTM, generally outperform others in terms of trace conservation, eigenvalue positivity, and prediction accuracy. However, we also observed that better physical constraint satisfaction does not always correlate with higher accuracy in dynamics prediction, highlighting important trade-offs when designing neural networks for quantum modeling.

Looking ahead, several promising directions can be pursued to further advance this approach. One such avenue is exploring the ability to learn from smaller subsystems and extrapolate to larger ones—for example, training on 
$N << M$ states and generalizing to 
$M > N$. Another valuable extension involves incorporating more complex and structured spectral densities to better capture realistic system-bath interactions. From an optimization perspective, replacing manually tuned loss weights with Lagrangian multipliers could offer a more principled and interpretable method for enforcing physical constraints, particularly within PINN-based architectures. Finally, the $\mathfrak{su}(n)$-based PINN framework employed in this work may also provide a natural foundation for tackling more advanced constraints, such as enforcing 
$n$-body positivity in systems with permutation symmetry—an important and currently underexplored area in quantum physics.

In summary, our Lie algebra-based ML framework offers a principled and scalable solution for modeling open quantum system dynamics. It inherently satisfies trace conservation, reduces reliance on complex loss terms, and facilitates more stable and efficient training. We believe this approach represents a significant step toward more accurate, generalizable, and physically consistent ML models for quantum dynamics.

\section{Acknowledgments}

A.U. acknowledges funding from the National Natural Science Foundation of China (No. W2433037) and Natural Science Foundation of Anhui Province (No. 2408085QA002). 

\section{Data availability}
The code and data supporting this work are available at \url{https://github.com/Arif-PhyChem/su_lie_algebra}.

\section{Competing interests}
The authors declare no competing interests.

\section*{References}
\bibliographystyle{vancouver}
\bibliography{main.bib,references.bib}

\end{document}


\title{Supporting Information for "Machine learning meets $\mathfrak{su}(n)$ Lie algebra: Enhancing quantum dynamics learning with exact trace conservation"}

\author{Arif Ullah}   
\email{arif@ahu.edu.cn}
\affiliation{School of Physics and Optoelectronic Engineering, Anhui University, Hefei, 230601, Anhui, China}  
\author{Jeremy O. Richardson}   
\email{jeremy.richardson@phys.chem.ethz.ch}
\affiliation{Department of Chemistry and Applied Biosciences, ETH Z\"urich, 8093 Z\"urich, Switzerland}  

\date{\today}

\begin{abstract}
\end{abstract}

\maketitle

\section{Details on data preparation and training}

\subsection{Choosing training trajectories}
To ensure a diverse and representative training set, we employ farthest-point sampling (FPS) to select trajectories from the full parameter space defined by the simulation parameters: characteristic frequency ($\gamma$), reorganization energy ($\lambda$), temperature ($T$) and $\beta$. Let each trajectory be represented as a point $x_i = (\gamma_i, \lambda_i, T_i)$ for FMO complex, or $x_i = (\gamma_i, \lambda_i, \beta_i)$ for SB model. Initially, one trajectory is chosen arbitrarily and then, for each unselected trajectory, we compute its minimum Euclidean distance to the set $S$ of already selected trajectories:

\begin{equation}
d(x_i, S) = \min_{x \in S} \|x_i - x\| \,.
\end{equation}

At each iteration, the trajectory with the largest distance is selected, i.e.,

\begin{equation}
x_{\text{next}} = \arg\max_{x_i \notin S} d(x_i, S) \,.
\end{equation}

This process maximizes the minimum distance between selected trajectories, ensuring that the sampled trajectories uniformly cover the parameter space. For each case of the SB model ($\varepsilon / \Delta =0$ and $1$) and the FMO complex (site-1 and 6), 400 trajectories were selected for training, and the remaining data were reserved for testing. Separate CNN-LSTM models were trained for the SB model and the FMO complex.

\subsection{Preparation of training data}
For the PUNN and PINN models, training data $(\mathbf{x}, \mathbf{y})$ is prepared using the OSTL scheme, where the full trajectory of the RDM, $\{\mathbf{\rho}_{\rm S}(t_0), \mathbf{\rho}_{\rm S}(t_1), \dots, \mathbf{\rho}_{\rm S}(t_M)\}$, serves as the target $\mathbf{y}$. At each time step, the RDM is flattened into a 1D vector by leveraging its Hermitian property: we retain all $n$ real-valued diagonal elements and the $\frac{n(n-1)}{2}$ unique upper off-diagonal elements, with each complex entry split into real and imaginary parts. The resulting vector follows the ordering:
\begin{align}
    [\rho_{11},\rho_{12}^{\text{real}}, \rho_{12}^{\text{imag}}, \rho_{13}^{\text{real}}, \rho_{13}^{\text{imag}}, \dots, \rho_{1n}^{\text{real}}, \rho_{1n}^{\text{imag}},\, \rho_{22},\, \rho_{23}^{\text{real}}, \rho_{23}^{\text{imag}}, \dots, \nonumber \\
    \rho_{2n}^{\text{real}}, \rho_{2n}^{\text{imag}},\, \dots,\, \rho_{(n-1)(n-1)},\, \rho_{(n-1)n}^{\text{real}}, \rho_{(n-1)n}^{\text{imag}},\, \rho_{nn}] \, , 
\end{align}
which results in $n^2$ entries per time step.

In contrast, for the $\mathfrak{su}(n)$-PUNN and $\mathfrak{su}(n)$-PINN models, the target $\mathbf{y}$ consists of the $n^2 - 1$ real-valued coefficients $a_i$ at each time step, obtained from the Lie algebra expansion.

The input $\mathbf{x}$ for all models is composed of the simulation parameters. For the SB model, these include $\varepsilon/\Delta$, $\Delta$, $\tilde{\lambda} = \lambda/\Delta$, $\tilde{\gamma} = \gamma/\Delta$, and $\tilde{\beta} = \beta \Delta$. To ensure consistent scaling across features, $\tilde{\lambda}$, $\tilde{\gamma}$, and $\tilde{\beta}$ are further normalized by their respective maximum values as $\tilde{\lambda}/\tilde{\lambda}_{\rm max}$, $\tilde{\gamma}/\tilde{\gamma}_{\rm max}$, and $\tilde{\beta}/\tilde{\beta}_{\rm max}$.

For the FMO complex, the simulation parameters include the initial excitation index, $\lambda$, $\gamma$, and $T$. The initial excitation is represented by a normalized label (e.g., 0.1 for an initial excitation on site-1, obtained by dividing the site index by 10), and the remaining parameters are normalized as $\lambda/\lambda_{\rm max}$, $\gamma/\gamma_{\rm max}$, and $T/T_{\rm max}$.

 \subsection{Training}
 For training we are using Keras Tensorflow and training it on system with CPUs of type Intel(R) Xeon(R) Gold 6230R CPU @ 2.10GHz.  Separate CNN-LSTM models were trained for the SB model and the FMO complex, each using an identical architecture. The structure of the CNN-LSTM model is given in Table~\ref{tab:cnn_lstm_model}. For comparison, we selected models with comparable training and validation losses as shown in Table~\ref{tab:cnn_lstm_losses}. 

\begin{supptable}[h]
\centering
\caption{CNN-LSTM model architecture and configuration where output size (OS) represents the length of the predicted trajectory.}
\label{tab:cnn_lstm_model}
\begin{tabular}{l l r}
\toprule
\multicolumn{3}{c}{\textbf{Model Configuration}} \\
\midrule
Batch Size & 64 & \\
Epochs & 3000 & \\
EarlyStopping Patience & 500 & \\
\midrule
\multicolumn{3}{c}{\textbf{Model Architecture}} \\
\midrule
\textbf{Layer (Type)} & \textbf{Output Shape} & \textbf{Param \#} \\
\midrule
Conv1D & (None, 3, 80) & 320 \\
Conv1D & (None, 3, 110) & 26,510 \\
Conv1D & (None, 3, 80) & 26,480 \\
MaxPooling1D & (None, 1, 80) & 0 \\
LSTM & (None, 112) & 86,464 \\
Dense & (None, 256) & 28,928 \\
Dense & (None, 128) & 32,896 \\
Dense & (None, 256) & 33,024 \\
Dense & (None, OS) & m = (256 $\times$ OS) + OS \\
\midrule
\multirow{2}{*}{Total Params} & \multicolumn{2}{r}{234622 + m} \\
\multirow{2}{*}{Trainable} & \multicolumn{2}{r}{234622 + m} \\
 \multirow{2}{*}{Non-trainable} & \multicolumn{2}{r}{0} \\
 \\
\bottomrule
\end{tabular}
\end{supptable}

\begin{supptable}[h]
\centering
\caption{Training and Validation Losses for CNN-LSTM Models}
\label{tab:cnn_lstm_losses}
\begin{tabular}{l l c c}
\toprule
\textbf{System} & \textbf{Model Type} & \textbf{Training Loss} & \textbf{Validation Loss} \\
\midrule
SB & PUNN & $1.544 \times 10^{-5}$ & $2.320 \times 10^{-5}$ \\
SB & $\mathfrak{su}(n)$-PUNN & $1.152 \times 10^{-5}$ & $3.191 \times 10^{-5}$ \\
SB & PINN & $1.611 \times 10^{-5}$ & $2.381 \times 10^{-5}$ \\
SB & $\mathfrak{su}(n)$-PINN & $1.382 \times 10^{-5}$ & $3.464 \times 10^{-5}$ \\
FMO & PUNN & $1.095 \times 10^{-5}$ & $1.245 \times 10^{-5}$ \\
FMO & $\mathfrak{su}(n)$-PUNN & $1.008 \times 10^{-5}$ & $2.274 \times 10^{-5}$ \\
FMO & PINN & $1.077 \times 10^{-5}$ & $1.292 \times 10^{-5}$ \\
FMO & $\mathfrak{su}(n)$-PINN & $1.059 \times 10^{-5}$ & $1.944 \times 10^{-5}$ \\
\bottomrule
\end{tabular}
\end{supptable}

\begin{suppfigure}[!thb]
    \centering
    \includegraphics[width=0.7\textwidth]{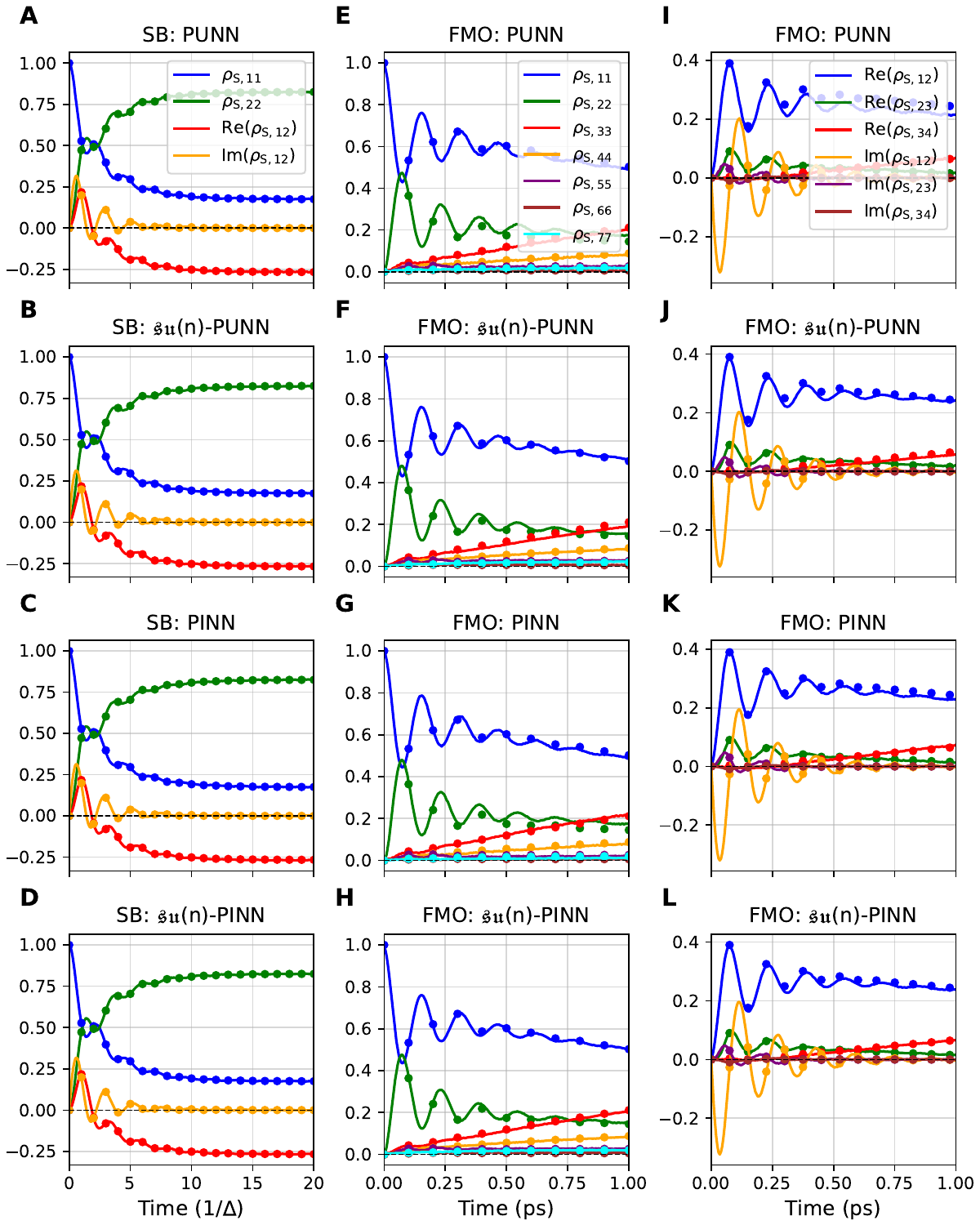}
    \caption{Evolution of the RDM's elements, including population and coherence terms, as predicted by PUNN (A, E, I), $\mathfrak{su}(n)$-PUNN (B, F, J), PINN (C, G, K), and $\mathfrak{su}(n)$-PINN (D, H, L) models. The first column shows the relaxation dynamics for the SB model (A-D), while the second (E-H) and third (I-L) columns highlight the diagonal (population) and selected off-diagonal (coherence) elements of the RDM for the FMO complex, respectively. Results are compared with reference dynamics, indicated by dots. Simulation parameters are provided in the caption of Fig.~2 (main text) and accuracy comparison is given in Table~I (main text).}
    \label{fig:dyn}
\end{suppfigure}

\begin{suppfigure}[!thb]
    \centering
    \includegraphics[width=0.7\textwidth]{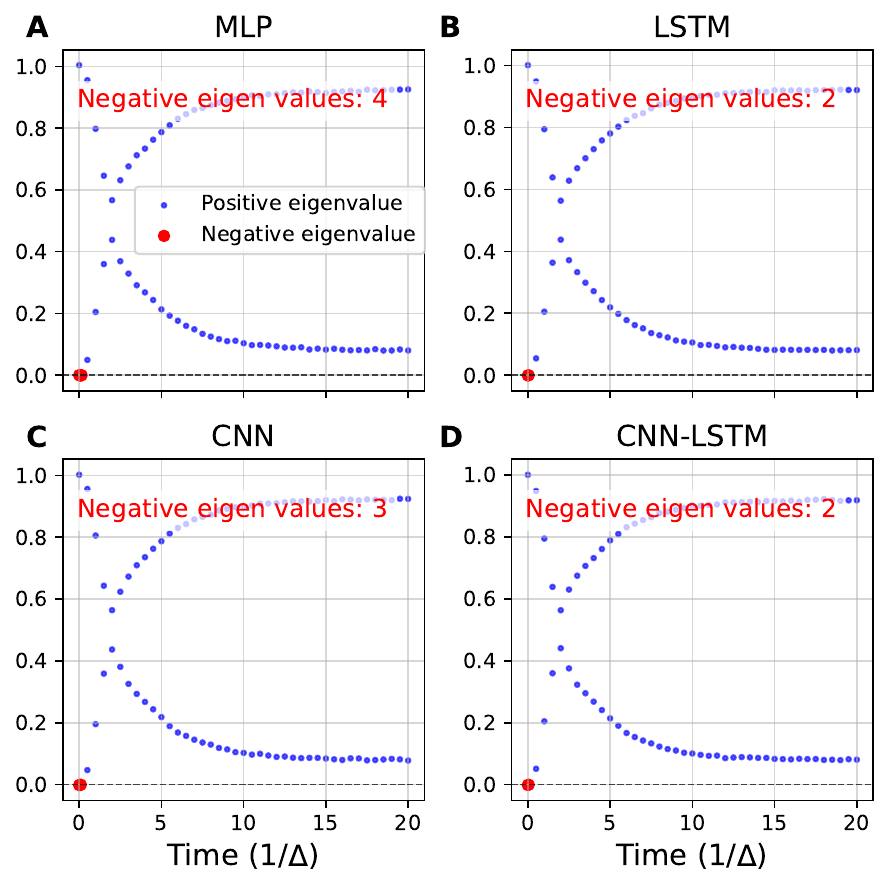}
    \caption{Evaluation of positive semi-definiteness and eigenvalue spectra of the predicted RDMs using the PUNN approach with (A) MLP, (B) LSTM, (C) CNN, and (D) CNN-LSTM architectures for the SB model. Sparse positive eigenvalues are shown for clarity, and the number of negative eigenvalues is reported for each model. Simulation parameters are consistent with those in Fig.~2 (main text).}
    \label{fig:compare_nn_sb_eig}
\end{suppfigure}

\begin{suppfigure}[!thb]
    \centering
    \includegraphics[width=0.7\textwidth, trim={0 {.05\textwidth} 0 0}, clip]{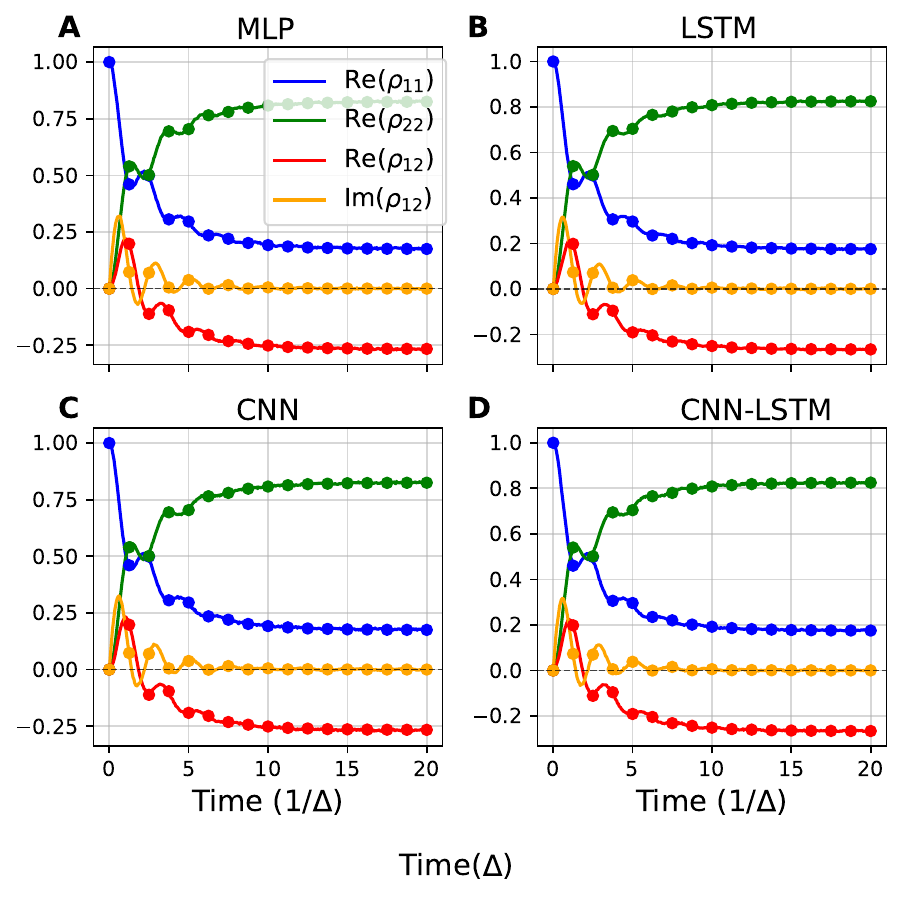}
    \caption{Time evolution of RDM elements—diagonal (population) and off-diagonal (coherence) components—predicted by (A) MLP, (B) LSTM, (C) CNN, and (D) CNN-LSTM architectures within the PUNN framework for the SB model. Reference dynamics is overlaid as dots. Simulation parameters are given in Fig.~2 (main text), and accuracy comparisons are summarized in Table~III (main text).}
    \label{fig:compare_nn_sb_dyn}
\end{suppfigure}